\documentclass
[prc,twocolumn,showpacs,preprintnumbers,amsmath,amssymb,superscriptaddress,floatfix,nofootinbib]{revtex4}
\usepackage{graphicx}
\usepackage{amsmath}
\usepackage{amsfonts}
\usepackage{slashed}
\usepackage{amssymb}
\newcommand{\smsm}[1]{\scriptscriptstyle{\scriptscriptstyle{#1}}}

\def\be{\begin{equation}}
\def\ee{\end{equation}}
\def\Be{\begin{eqnarray}}
\def\Ee{\end{eqnarray}}
\def\ba{\begin{array}}
\def\ea{\end{array}}
  \def\CL{{\cal L}}
\def\CM{{\cal M}}

\begin{document}

\title{Studying triangle singularity through spin observables}

\author{Ke Wang} \author{Shao-Fei Chen}\author{Bo-Chao Liu} \email{liubc@xjtu.edu.cn}

\affiliation{MOE Key Laboratory for Nonequilibrium Synthesis and Modulation of Condensed Matter, School
of Physics, Xi’an Jiaotong University, Xi’an 710049, China.}
\affiliation{
Institute of Theoretical Physics, Xi’an Jiaotong University, Xi’an 710049, China.
}

\begin{abstract}

In this work, we study the spin density matrix element $\rho_{00}$
of the $\phi$ in the decay $J/\psi \rightarrow \eta \pi \phi$. In
previous studies, a band around 1.4 GeV on the $\pi^0\phi$ distribution
in Dalitz plot was reported by the BESIII Collaboration. This structure
may be caused by the production of a resonance or the triangle
singularity mechanism. We find that the predictions of the spin density
matrix elements of the final $\phi$ based on these mechanisms show
distinct features. Thus the measurement of the spin density matrix
elements of the $\phi$ in this reaction may offer an alternative way
to study the triangle singularity and to clarify the reaction mechanisms,
i.e. resonance production or kinematic effects. This work also shows the
potential of spin observables in studying kinematic singularities.
\end{abstract}
\maketitle

\section{INTRODUCTION}
The studies on the hadron spectrum offer the platform to test our
knowledge of quantum chromodynamics(QCD) in the nonperturbative
regime, which is important for understanding the strong
interactions. In recent years, owing to a large amount of new
experimental results on particle reactions and resonances there have
been significant progress in the study of the hadron spectrum. A
large number of new states were found, which usually show as peaks
or dips in the invariant mass spectrum of final particles. While, a
peak in the invariant mass spectrum is not necessarily caused by a
resonance. It also can be produced by kinematic effects. In fact,
some of the new states are interpreted as threshold cusps and/or
triangle singularities\cite{a1:1420a,a1:1420b,f2:1810,X3872,L1405,
L1670,Pc,guoreview1,guoreview2}. Since these kinematic effects may
show similar features as resonances, it is then important to find
some ways to distinguish the kinematic singularities from genuine
resonances.

Besides the interests in clarifying the nature of the observed structures in experiments, triangle singularity(TS) mechanism may also play an essential role in understanding some important puzzles. Some remarkable examples can be found in relevant studies in $J/\psi$ decays. In 2012, BESIII Collaboration reported the observation of abnormally large isospin-breaking effects in $J/\psi\to\gamma\eta(1405/1475)\to\gamma+3\pi$\cite{BESIII:2012aa}, which, however, could be understood by considering the important roles of the TS mechanism via the intermediate $K^*\bar K + c.c.$ rescatterings in this decay. Furthermore, it was also argued that the TS mechanism could be crucial for understanding the nature of $\eta$ resonances\cite{Du:2019idk,Cheng:2021nal} and the productions and decays of light axial vector mesons\cite{zhao2021} in $J/\psi$ decays. Even though the TS mechanism may play important roles in the physical processes mentioned above, further experimental evidences are still needed to identify its contribution. Up to now, most studies on the TS mechanism mainly concentrate on its effects in the invariant mass spectrum. In this work, we hope to show that the TS mechanism may also cause significant spin effects and in some cases spin observables
are helpful for identifying its contributions.
Here we will concentrate on the reaction $J/\psi \rightarrow \eta \pi
\phi$. In Ref.\cite{GUO}, it was argued that if considering the
contributions from a set of $K^*K\bar K$ triangle
diagrams[Fig.\ref{feynfig}(a)] a peak around 1.4 GeV in
the $\pi^0\phi$ invariant mass distribution can be produced in the
$J/\psi \rightarrow \eta \pi \phi$ reaction, which fits well with the recent
measurement on this reaction by BESIII Collaboration\cite{bes}. Of
course, the peak observed by experiment can also be interpreted by
considering the production of a resonance[Fig.\ref{feynfig}(c)]. In
Ref.\cite{GUO}, the authors suggested that by checking whether the
structure around 1.4 GeV persists for the $K^+K^-$ invariant mass
away from the $\phi$ mass region one could distinguish these two
models. Later, the authors in Ref.\cite{zhao2021} argued that the decay
could also proceed with the production of $h_1(1415)$ at first and then $h_1(1415)$
decaying to $\pi\phi$ through the same triangle diagram[Fig.\ref{feynfig}(b)]. Both the models in Refs.\cite{GUO,zhao2021} concern the TS mechanism, and we
call them the TS models to distinguish from the resonance model where a resonance is produced and decays without significant 
rescattering effects. In this work, we will show that it is possible to
distinguish these two kinds of models by measuring the spin density matrix elements(SDMEs) of the final $\phi$.

This paper is organized as follows. In Sec. II, the theoretical
framework and ingredients are presented. In Sec. III, the numerical
results are presented with some discussions. Finally, the paper ends
with a short summary in Sec. IV.
\section{MODEL AND FORMALISM}
\begin{figure}[htbp]
\begin{center}
\includegraphics[scale=0.5]{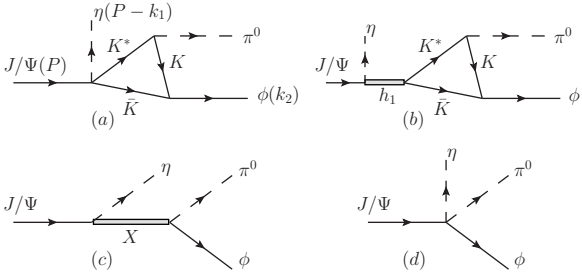}
\caption{Feynman diagrams for the $J/\psi \rightarrow \eta \pi \phi$
reaction.} \label{feynfig}
\end{center}
\end{figure}

As mentioned in Sec.I, a peak structure being around 1.4 GeV in the
$\pi\phi$ invariant mass spectrum was found in the reaction $J/\psi
\rightarrow \eta \pi \phi$\cite{bes}. This peak may be caused by a resonance(resonance model), e.g. C(1480)\cite{g13,C1480}/$h_1(1415)$\cite{pdg,bes:h1415,zhao},
or triangle diagrams involving $K^*K\bar K$ intermediate
states(TS model) as discussed in Refs.\cite{GUO,zhao2021}. The Feynman diagrams for the
TS and resonance models can be depicted
in Fig.\ref{feynfig}. To calculate these Feynman diagrams, the Lagrangian
densities for the involved vertices are needed. For the triangle diagrams,
we basically follow the formalism in Ref.\cite{GUO}, where the
triangle diagrams were calculated. As shown in Ref.\cite{GUO}, the
amplitudes for Fig.\ref{feynfig}(a) can be presented as
\begin{widetext}
\begin{eqnarray}
\CM_{i} &=& ig \epsilon^\mu_{J/\psi}
\epsilon^{*\nu}_\phi \int\frac{d^4q}{(2\pi)^4}
\frac{[-g_{\mu\lambda}+(q+k_1)_\mu(q+k_1)_\lambda/m^2_{K^*_{id}}](q+2k_2-k_1)^\lambda(2q+k_2)\nu}
{(q^2-m^2_{K_{id}}+i\epsilon)[(q+k_1)^2-m^2_{K^*_{id}}+i\epsilon][(q+k_2)^2-m^2_{K_{id}}+i\epsilon]}
\equiv i g \epsilon^\mu_{J/\psi}
\epsilon^{*\nu}_\phi \CM^{i}_{\mu\nu},\nonumber\\
\label{ampdef}
\end{eqnarray}
\end{widetext}
where the index $i=C(N)$ denotes the amplitudes corresponding to the
process with the charged (neutral) intermediate particles and g is a constant. The
concrete expressions of $\CM^{i}_{\mu\nu}$ and meanings of parameters can be found in Ref.\cite{GUO}. By summing the charged and neutral loop
amplitudes, with including appropriate coefficients, the total
amplitude can be presented as
\begin{eqnarray}
\CM &=& 2(\CM_C-\CM_N)\label{tamp}
\end{eqnarray}

For the amplitude of Fig.\ref{feynfig}(b), where $h_1(1415)$ is produced at first, the total amplitude can also be written as Eq.(\ref{tamp})
but with the $\CM_{i=C(N)}$ being replaced by\cite{zhao2021,zhao}
\begin{eqnarray}
\CM_{i} \equiv i \frac{g' \epsilon_{J/\psi,\rho} \left(-g^{\rho\mu}
+ k^\rho_1 k^\mu_1 / k^2_1 \right) \epsilon^{*\nu}_\phi
\CM^{i}_{\mu\nu}}{k^2_1 - m^2_{h_1} + i m_{h1} \Gamma_{h_1}},
\label{diff}\end{eqnarray}
where the $\CM^{i}_{\mu\nu}$ is defined in Eq.(\ref{ampdef}) and $g'$ represents the coupling constant. 

To calculate the tree diagram in the case of the C(1480) production,
we need to consider the process represented
by Fig.\ref{feynfig}(c) with taking X as C(1480)($I=1$,$J^{PC}
=1^{--}$)\cite{g13,C1480}. The effective Lagrangians for the $J/\psi X\eta$
and $X\pi\phi$ vertices are adopted as\cite{VVP}

\begin{eqnarray}
\CL_{VVP}=g_{\smsm{V}}\varepsilon^{\mu\nu\alpha\beta}\partial_\mu
V_\nu
\partial_\alpha V_\beta P,
\end{eqnarray}
where V denotes the field of a vector meson($J/\psi$ or $\phi$), and
P denotes the field of a pseudoscalar meson($\eta$ or $\pi$). Note
that isospin invariance needs not to be considered for the
$J/\psi X\eta$ or $X\pi\phi$ vertex depending on the isospin of $X$ being 1 or 0 respectively, because isospin conservation is violated in this decay. The amplitude for this tree diagram can then be
obtained as:

\begin{eqnarray}
-i\CM_{C(1480)}=g_{\smsm{C}}\varepsilon_{\mu\nu\alpha\beta} p_{\phi}^\mu \phi^{*\nu}
p_C^\alpha G_1^{\beta b}(p_C) \varepsilon_{abcd} p_C^a p_\psi^c
\psi^d,
\end{eqnarray}
where $g_{\smsm{C}}$ represents the product of coupling constants in this
process, and the $G_1^{\mu \nu}$ is taken as:
\begin{eqnarray}
G_1^{\mu \nu}(p_X)=\frac{-g^{\mu \nu}+\frac{p_X^\mu p_X^\nu}
{p_X^2}}{p_X^2-m_X^2+i m_X \Gamma_X}.
\end{eqnarray}
For the mass and
width of the C(1480), we adopt $m_{\smsm{C(1480)}}=1480$ and $\Gamma_{\smsm{C(1480)}}=130$ MeV
\cite{g13,C1480}.

For the case that the intermediate state X is the $h_1(1415)$, the Feynman diagram for the process can also be
presented by Fig.\ref{feynfig}(c) with taking X as $h_1(1415)$.
Since $h_1$ has quantum numbers $I=0$ and $J^{PC}=1^{+-}$, the effective
Lagrangian for the $J/\psi h_1\eta$ and $h_1\phi\pi$ vertices can be
written as\cite{AVP1,AVP2}

\begin{eqnarray}
\CL_{AVP} &=& g_{\smsm{A}}(\CL_a
cos\theta+\CL_bsin\theta),\label{ma2}\\
\CL_a  &=& A^\mu(\partial_\mu V_\nu-\partial_\nu V_\mu)\partial^\nu P,\\
\CL_b  &=&\partial^\mu A^\nu(\partial_\mu V_\nu-\partial_\nu V_\mu)
P,
\end{eqnarray}
where V denotes $J/\psi$ or $\phi$ field, P denotes $\pi$ or $\eta$ field and A
represents $h_1$ field. The $g_A$ and $\theta$ represent the coupling constant and mixing angle.

Then the corresponding amplitude for Fig.\ref{feynfig}(c) can be
obtained as
\begin{eqnarray}
-i\CM_h &=& g_h  G_1^{\mu a}(p_{h_1})\cdot\nonumber\\&&
(p_{\phi\mu}\phi_\nu-p_{\phi\nu}\phi_\mu) (p_\pi^\nu cos \theta_\pi
+p_{h_1}^\nu sin \theta_\pi)\cdot\nonumber\\&& (p_{\psi
a}\psi_b-p_{\psi b} \psi_a)(p_\eta^b cos \theta_\eta -p_{h_1}^b sin
\theta_\eta),
\end{eqnarray}
where $\theta_\eta$ and $\theta_\pi$ represent the mixing angles in
the Lagrangians of the $J/\psi h_1\eta$ and $h_1\phi\pi$ vertices,
respectively. The mass and width
of $h_1$ can be taken from PDG book\cite{pdg} as $m_{h_1}=1416$
and $\Gamma_{h_1}=90$ MeV.

\section{The spin density matrix elements} With the amplitudes
given above, the spin density matrix elements of the final $\phi$ in
the $J/\psi\to \phi\eta\pi$ reaction can be calculated. Since the
initial $J/\psi$ considered in this work is produced in $e^+e^-$
collisions, we choose the polarization axis of $J/\psi$ along
z-axis, which is defined as the beam direction of $e^+$ or $e^-$. In
this case, the magnetic quantum numbers of the
$J/\psi$ only takes the values $m=\pm 1$\cite{jpsi0}. For the final $\phi$, we
shall consider its helicity states, i.e. choosing its polarization
axis along its momentum direction, in the c.m. frame of the $\pi^0
\phi$ system. The spin density matrix element $\rho_{00}$ of the $\phi$(denoted as $\rho_{00}^\phi$) as a function of the $\pi^0\phi$ invariant
mass in the $\pi^0 \phi$ rest frame is defined as

\begin{eqnarray}
\rho_{00}^\phi(m_{\pi^0\phi}) = \frac{\int {\rm d}
\Omega_\eta {\rm d} \Omega_{\pi^0} \sum\limits_{m} \CM_{m,\lambda=0}
\CM^*_{m,\lambda'=0}}{\int{\rm d} \Omega_\eta {\rm d}
\Omega_{\pi^0}\sum\limits_{m,\lambda^{''}}|\CM_{m,\lambda^{''}}|^2},
\label{def}\end{eqnarray}
where $m(=\pm 1)$ represents the z-component of the total angular
momentum of $J/\psi$ and $\lambda$, $\lambda'$ and $\lambda^{''}$
are the helicities of the final $\phi$. The $\rho_{00}^\phi$ can be
extracted from the angular distribution of $K$ or $\bar K$ in
$\phi\to K\bar K$ through\cite{AD1,AD2,AD3}
\Be W(cos\theta)\sim\frac{3}{2}[\rho_{00}^\phi cos^2\theta+\frac{1}{2}(1-\rho_{00}^\phi)sin^2\theta], \Ee
where $\theta$ is defined in the conventions of the helicity system.

\section{RESULT AND DISCUSSION}

As we know, both the kinematic singularity and genuine
resonance state can result in a structure in the invariant mass
spectrum. Therefore, it is interesting and important to find some other observables to distinguish these two mechanisms. In this section, we shall study the dependence of the SDME $\rho_{00}^\phi$ on the invariant mass spectrum $m_{\pi\phi}$ considering different mechanisms, and then discuss the possibility of distinguishing various mechanisms using this observable. For the purpose of this work, it is helpful to firstly study the features of the $\rho_{00}^\phi$ induced by the mechanisms shown in Fig.\ref{feynfig} individually. In doing so, the
coupling constants are irrelevant.
Therefore, we just set all the coupling constants as 1, and the possible effects from
background contribution will be discussed later. For reader's convenience, the
parameters for the resonance models are collected and listed in Table 1.

\begin{figure}[htbp]
\begin{center}
\includegraphics[scale=0.4]{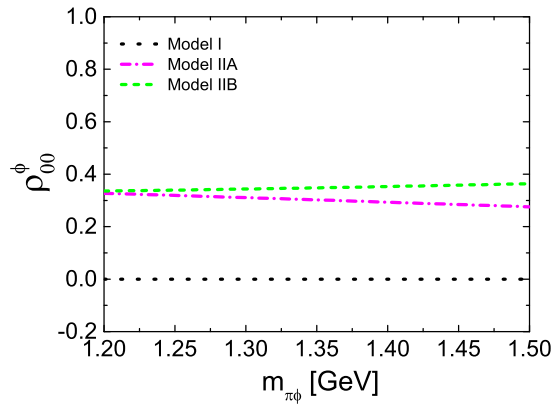}
\caption{The obtained $\rho_{00}^\phi$ for the tree diagrams. The
dotted(black), short-dash-dotted(magenta) and short-dashed(green)
lines represent the results of the $C(1480)$ production(Model I), $h_1(1415)$ production with taking $\theta_\pi=\pi/4$ (Model IIA) and $\theta_\pi=3\pi/4$(Model IIB), respectively. }
\label{rho1f}
\end{center}
\end{figure}

In the case of the resonance production process with taking
$X= C(1480)$ (Model I), the SDME $\rho_{00}^\phi$ is 0 as shown by
the black dotted line in Fig.\ref{rho1f}. This means that the final
$\phi$ meson can only be in the helicity states with $\lambda=\pm1$.
This results from
the properties of the $C\pi \phi$(V-V-P coupling) vertex, which is
discussed in detail in Appendix~\ref{VP}.

In the case that X is the $h_1(1415)$, to calculate the amplitude we
have to fix the values of the mixing angles $\theta_\eta$ and
$\theta_\pi$ in the Lagrangians first. In general, the mixing angles
should be determined by fitting the experimental data. Unfortunately,
up to now the information about these angles is still absent.
In our calculations, we find that the value of the $\rho_{00}^\phi$
is independent of the $\theta_\eta$. On the other hand, the value of
$\theta_\pi$ is relevant to the $\rho_{00}^\phi$. If we take
$\theta_\pi=\pi/4$ (Model IIA), which leads to a mixing of S-wave and
D-wave couplings, the value of the $\rho_{00}^\phi$ is about 0.30(the
magenta short-dash-dotted line in Fig.\ref{rho1f}). If we take
$\theta_\pi=3\pi/4$ (Model IIB), the $\CL_{h_1\phi\pi}$ describes the
S-wave coupling, and the value of $\rho_{00}^\phi$ is about 0.35(the
green short-dashed line in Fig.\ref{rho1f}). In a general case, the
mixing angles can be arbitrary and control the relative importance
of the S-wave and D-wave couplings. However, we find that the D-wave
contribution is suppressed compared to the S-wave contribution, and
the mixing angle only has minor effects on the value of the $\rho_{00}^\phi$
except for taking $\theta_\pi$ at some special value.\footnote{In fact,
we find that when the $h_1\pi\phi$ coupling has the form $h_1^\mu\partial_\mu \phi_\nu\partial^\nu \pi$
the $\rho_{00}^\phi$ can approach 1. But this only happens
in a very small parameter space of the $\theta_\pi$, otherwise the
dependence of the $\rho_{00}^\phi$ on the $\theta_\pi$ is rather weak.
So in this work we ignore the possibility that the $h_1\pi\phi$ vertex
has this special coupling, which can certainly be verified by future
studies on the $h_1\pi\phi$ coupling.} Therefore, in the $h_1(1415)$
production case we conclude that the $\rho_{00}^\phi$ tends to be a
relatively small value.
\begin{table}[htbp]
\caption{Parameters for resonance models.}
\begin{tabular}{cccccc}
  \hline\hline
  Model & Resonance & $J^{PC}$ &  Mass(GeV)  &  Width(GeV)   & $\theta_\pi$  \\
  \hline
  I    &  $C(1480)$  & $1^{--}$ & 1.480 &  0.13  & -           \\
  IIA  &  $h_1(1415)$ & $1^{+-}$ & 1.416 &  0.09  & $\pi/4$           \\
  IIB  &  $h_1(1415)$ & $1^{+-}$ & 1.416 &  0.09  & $3\pi/4$           \\
  \hline\hline
\end{tabular}
\label{tab1}
\end{table}

\begin{figure}[htbp]
\begin{center}
\includegraphics[scale=0.4]{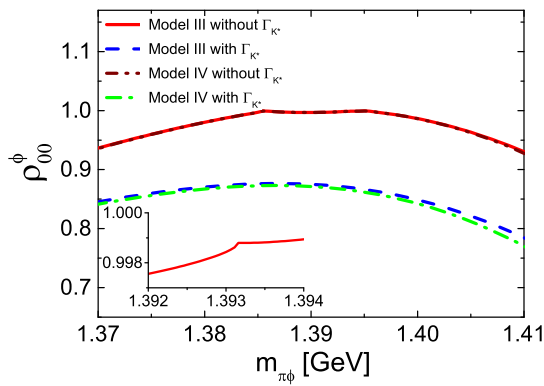}
\caption{The obtained $\rho^\phi_{00}$ for Model III and Model IV near the TS with and without considering the width of $K^*$.} \label{rhoTS1}
\end{center}
\end{figure}

In Refs.~\cite{GUO,zhao2021}, the authors have analyzed the
$J/\psi\to\eta\pi^0\phi$ reaction by considering the triangle
diagrams Fig.\ref{feynfig}(a)(Model III)\cite{GUO} or Fig.\ref{feynfig}(b)(Model IV)\cite{zhao2021} and shown that those
diagrams can cause a peak around 1.4 GeV in the $\pi^0\phi$
invariant mass distribution. In fact, there are two kinds of
singularities which are relevant~\cite{ts}. One is the normal two-body
threshold cusp (TBTC), and the other is the triangle singularity.
Using the parameters of the particles from the PDG book~\cite{pdg},
the TBTC and TS for the diagrams with the charged intermediate
states are located at 1.3853 and
1.3857 GeV, respectively. And for diagrams with the neutral
intermediate states they are located
at 1.3931 and 1.3952 GeV, respectively. The SDME $\rho_{00}^\phi$
for the triangle diagrams(Model III and IV) near the TS has been studied with
or without considering the width of the $K^*$ in the loop
in Fig.\ref{rhoTS1}. In all TS models, the $\rho_{00}^\phi$
is always larger than 0.75. When neglecting the $K^*$ width, there are
two peaks in the distribution of $\rho_{00}^\phi$ versus the $\pi^0\phi$
invariant mass. At the peaks, the $\rho_{00}^\phi$ approaches 1.
After including the $K^*$ width effects, the distribution of
$\rho_{00}^\phi$ only has one relatively wide peak. At the same
time, the value of $\rho_{00}^\phi$  will decline to $0.77\sim0.88$.

The relatively large value of $\rho_{00}^\phi$ induced by the triangle
diagrams may be ascribed to the properties of the $K\bar{K}\phi$ vertex.
Taking Fig.\ref{feynfig}(a) as an example, if the three-momenta of $K$
and $\phi$ are collinear, a large value of $\rho_{00}^\phi$ will be
obtained as discussed in Appendix \ref{VP}. Since the considered invariant
mass $m_{\pi\phi}$ in Fig.\ref{rhoTS1} is near the $\bar KK^*$ threshold, the
magnitude of the three momentum of the $\bar K$ is close to zero when the
intermediate states $K^*$ and $\bar K$ in the loop are on shell. In this case,
the three-momenta of $K$ and $\phi$ are approximately collinear. Since
it is expected that when the intermediate states are on shell the amplitude
will get a relatively large value, the case discussed above gives the main
contribution in the loop integral\cite{zhao}. Thus the density matrix element
$\rho_{00}^\phi$ tends to have a large value. This property also leads to
the cusp at TBTC (see the inset plot in Fig.\ref{rhoTS1}). Based on the same logic, without considering the $K^*$'s width the
$\rho_{00}^\phi$ should approach 1 at TS, where the three intermediate
particles are on shell and moving collinearly~\cite{ts}.

\begin{figure}[htbp]
\begin{center}
\includegraphics[scale=0.4]{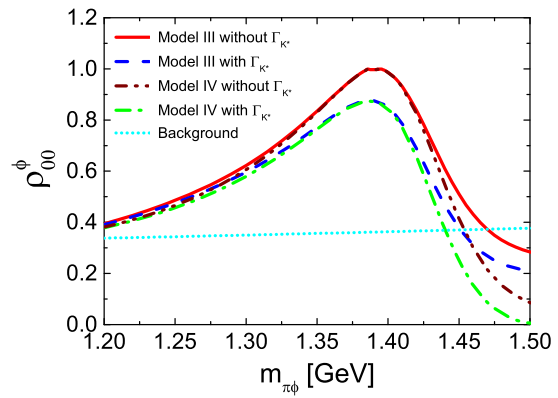}
\caption{The obtained $\rho_{00}^\phi$ for Model III and Model IV in a
wider range of $m_{\pi\phi}$ with and without considering the width of $K^*$.The results for background contribution is also shown for comparison.} \label{rhoTS2}
\end{center}
\end{figure}

On the other hand, when the value of $m_{\pi\phi}$ moves away from the
$\bar KK^*$ threshold, the collinear condition does not hold anymore.
Therefore, the value of $\rho_{00}^\phi$ decreases as $m_{\pi\phi}$
departing from the locations of TS and TBTC. In Fig.\ref{rhoTS2}, we
show the SDME $\rho_{00}^\phi$ in a wider range of $m_{\pi\phi}$.
No matter whether the width of $K^*$ is considered, the $\rho_{00}^\phi$
distribution shows a wide peak, which is peaked at around
$m_{\pi\phi}=1.39$ GeV. Comparing with the results of the resonance
models in Fig.\ref{rho1f}, it is clear that the
$\rho_{00}^\phi$ show distinct features for the various models which
all can explain the peak at around 1.4 GeV in the $\pi^0\phi$ invariant
mass distribution. Therefore, the $\rho_{00}^\phi$ has the potential
to clarify whether the structure in the invariant mass spectrum is
caused by genuine resonance or by TS mechanism.

Concerning the $\rho_{00}^\phi$ focused in this work, the main discrepancies between Models III and IV appear at higher $m_{\pi\phi}$. In fact, if one looks at the structures of the amplitudes of the two models, the main difference originates from the term $\frac{p^\mu p^\nu}{p^2}$ in the propagator of $h_1$ in Model IV(see Eq. (\ref{diff})). Note that the denominator of the $h_1$ propagator is canceled in the calculations of the $\rho_{00}^\phi$ (see Eq. (\ref{def})). The presence of the $\frac{p^\mu p^\nu}{p^2}$ term guarantees that the total angular momentum of the $\pi\phi$ system is 1. While, at lower $m_{\pi\phi}$ the dominance of the S-wave component of the final $\pi\phi$ and intermediate $\bar KK^*$ systems automatically enforce that the total angular momentum of the $\pi\phi$ system is 1. Therefore, the two models give similar results. At higher $m_{\pi\phi}$, the $\frac{p^\mu p^\nu}{p^2}$ term starts to play a more important role, and the difference between the two models become evident. Numerically, if one calculates the contributions from the $-g_{\mu\nu}$ and $\frac{p^\mu p^\nu}{p^2}$ terms individually, it can be found that the $-g_{\mu\nu}$ term gives the dominant contribution at lower $m_{\pi\phi}$. At the region above the peak, the contribution of the $-g_{\mu\nu}$ term decreases more quickly than that of the $\frac{p^\mu p^\nu}{p^2}$ term, and then the $\frac{p^\mu p^\nu}{p^2}$ term becomes more important at higher $m_{\pi\phi}$. It is also interesting to note that because of the destructive effects between these two terms the $\rho_{00}^\phi$ drops faster in Model IV than in Model III. In particular, in Model IV with taking into account the $K^*$'s width the $\rho_{00}^\phi$ may approach 0 at about $m_{\pi\phi}$=1.50 GeV. 

\begin{figure}[htbp]
\begin{center}
\includegraphics[scale=0.4]{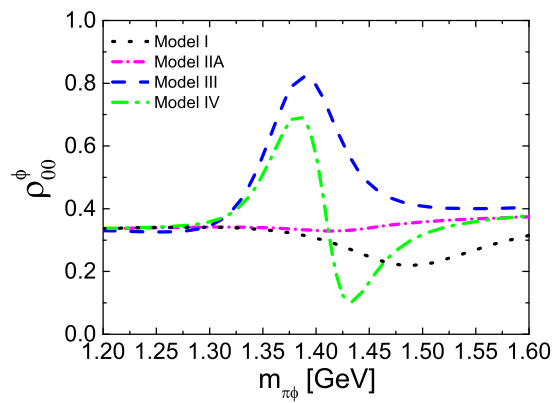}
\caption{The calculated $\rho_{00}^\phi$ for various models with including
the background contribution. For Model III and Model IV, the finite width of $K^*$ has been taken into account.}\label{all}
\end{center}
\end{figure}

Finally, to compare with experimental data, it is also necessary
to estimate possible effects from the background contributions.
Possible resonance contributions in the $\pi^0\eta$ channel, such as the $a_0(980)$'s contribution, are not considered, since they can, in principle, be eliminated by a kinematic cut on the $\pi^0\eta$ invariant mass. In this work, the background contribution is modeled by a contact
term (Fig.\ref{feynfig}d), for which we adopt the Lagrangian
density\cite{GUO,Meissner:2000bc},
\begin{eqnarray}
\CL_{\Psi\eta\pi\phi}=g_{ct} \Psi^\mu \phi_\mu\pi\eta.
\end{eqnarray}
Since the relative strength of the background contribution is not
presented in the experimental paper, here we adjust the coupling
constant $g_{ct}$ to make the background contribution have the same
magnitude as that of the resonance contribution or the triangle
diagrams at the peak position in the invariant mass spectrum.
In this way, the $\rho_{00}^\phi$ with including background
contribution is calculated for various models and shown in
Fig.\ref{all}\footnote{In Fig.\ref{all}, we only show the
results corresponding to the constructive interference case.
For the triangle diagrams, we do not consider the destructive
interference case, since in this case we can not get a peak
structure in the invariant mass spectrum, which is conflict
with the experimental observation. For the resonance production
process, we find the interference effects are insignificant and
the results are similar in both constructive and destructive cases.}.
In the region where the background term dominates the reaction,
the value of $\rho_{00}^\phi$ approaches 0.33 corresponding to
the pure background contribution (see short-dotted line in Fig.\ref{rhoTS2}) for all models. At the peak
position, the resonance or triangle diagram contribution has a
similar strength as the background contribution as we suppose.
For TS models, we find the $\rho_{00}^\phi$
is slightly reduced. For the $h_1(1415)$ production process(Model II), since
the resonance production contribution and background term individually
leads to a similar $\rho_{00}^\phi$, we find the inclusion of the
background contribution does not significantly change the
$\rho_{00}^\phi$. While for the C(1480) production process(Model I), the
value of the $\rho_{00}^\phi$ is determined by a mix of the background
and resonance contribution. In this case the $\rho_{00}^\phi$ always
lies in a range between the values determined by the C(1480) contribution
and background contribution solely, i.e. in a range from 0 to 0.33.
Therefore, we find that although the inclusion of the background
contribution could change the line shapes of $\rho_{00}^\phi$ for
different models, the main difference between the TS
models and resonance models remains and can be used
to distinguish various mechanisms. In particular, the discrepancies between Model III and IV at higher $m_{\pi\phi}$ still exist and may offer the opportunity to distinguish these two models. However, since the contributions due to the TS mechanism become smaller at higher $m_{\pi\phi}$, it may be challenging to capture such discrepancies.   

\section{SUMMARY}
In this work, we study the spin density matrix element $\rho^\phi_{00}$
of the $\phi$ in the reaction $J/\psi\to \eta\pi\phi$. We find that the
$\rho^\phi_{00}$ shows distinct features when considering different
reaction mechanisms, i.e. the production of a resonance or TS mechanism. According to our calculation results, it is found
that the special kinematic conditions required by kinematic singularities and the properties of the involved vertex functions in the loop
results in an enhancement of the $\rho^\phi_{00}$ near the TS. If the
TS mechanism indeed plays an important role, we expect that  the
$\rho^\phi_{00}$ should take a relatively large value and have a peak
versus the invariant mass $m_{\pi\phi}$ near the TS, which is absent
for the resonance models. Therefore, by exploring the
$\rho^\phi_{00}$ in this reaction it is possible to distinguish the
various models and offer an alternative way to study the triangle
singularity. Untill now there is still no clear experimental evidence
identifying the contribution of triangle singularity, it is then helpful to develop some new methods to distinguish various mechanisms. Although in this work we concentrate on the $J/\psi\to \eta\pi\phi$ reaction, it should be noted that the spin effects caused by the TS mechanism are general and can be exploited in other processes where the TS mechanism plays an important role and the spin states of final particles can be measured.

\section*{Acknowledgments}
We acknowledge Professor Feng-Kun Guo for his useful comments and suggestions. This work is partly supported by the National Natural Science
Foundation of China under Grants No.U1832160 and No.11375137, the
Natural Science Foundation of Shaanxi Province under Grant
No.2019JM-025, and the Fundamental Research Funds for the Central
Universities.
\begin{appendix}\section{POLARIZATION VECTORS OF VECTOR MESON}\label{POV}
Taking the polarization axis along the z-axis, the polarization vectors
for vector meson at rest are \Be
\varepsilon_{+1}(\vec{p}=0)=&&\frac{-1}{\sqrt{2}}\left
(\begin{array}{c} 0 \\ 1\\i\\0 \end{array} \right ),
\varepsilon_{-1}(\vec{p}=0)=\frac{1}{\sqrt{2}}\left (\begin{array}{c} 0
\\ 1\\-i\\0 \end{array} \right ),\nonumber\\&&\varepsilon_{0}(\vec{p}=0)=\left
(\begin{array}{c} 0 \\ 0\\0\\1 \end{array} \right ) \Ee
 For vector
meson polarized along the direction with spherical angles
($\theta,\phi$) in its rest frame, the polarization vectors are
obtained through
\begin{eqnarray}
\varepsilon'_\lambda (\theta,\phi) &=& \sum\limits_{M}D_{M\lambda}
(\phi, \theta, -\phi) \varepsilon_M \\\nonumber
&=&e^{-i\phi\cdot1}d^1_{1\lambda}(\theta)e^{i\phi\cdot\lambda}\varepsilon_1
+e^{-i\phi\cdot0}d^1_{0\lambda}(\theta)e^{i\phi\cdot\lambda}\varepsilon_0
\\\nonumber
&&+e^{-i\phi\cdot(-1)}d^1_{-1\lambda}(\theta)e^{i\phi\cdot\lambda}\varepsilon_{-1}.
\end{eqnarray}
Then we can get

\begin{eqnarray}
\varepsilon '_{+1}(\vec{p}=0)&=&\frac{-1}{\sqrt{2}} \left(
  \begin{array}{c}
    0 \\
    \cos^2\frac{\theta}{2}-e^{2i\phi}\sin^2\frac{\theta}{2} \\
   i(\cos^2\frac{\theta}{2}+e^{2i\phi}\sin^2\frac{\theta}{2}) \\
   -e^{i\phi}\sin\theta
  \end{array}
\right)\\
 \varepsilon '_{-1}(\vec{p}=0)&=&\frac{-1}{\sqrt{2}}
\left(
  \begin{array}{c}
    0 \\
    -\cos^2\frac{\theta}{2}+e^{-2i\phi}\sin^2\frac{\theta}{2} \\
   i(\cos^2\frac{\theta}{2}+e^{-2i\phi}\sin^2\frac{\theta}{2}) \\
   e^{-i\phi}\sin\theta
  \end{array}
\right) \\
 \varepsilon '_{0}(\vec{p}=0)&=&\left(\begin{array}{c}
     0 \\
     \sin\theta\cos\phi\\
     \sin\theta\sin\phi\\
     \cos\theta
     \end{array}\right)
\end{eqnarray}

For taking a vector meson from rest to momentum p, the corresponding
lorentz transformation matrix is defined as \be \left
(\begin{array}{c} p^0 \\ p^j \end{array} \right )=\frac{1}{M_V}\left
( \begin{array}{cc} p^0 & p^i \\ p^j &
\frac{p^jp^i}{p^0+M_V}+\delta^{ji}M_V\end{array} \right )  \left
(\begin{array}{c} M_V \\ 0 \end{array} \right ), \ee where $M_V$
represents the mass of the vector meson. So for a vector
meson moving at the momentum p and polarized along $(\theta,\phi)$,
the polarization vectors are
\begin{eqnarray}
\varepsilon_\lambda(p)=\frac{1}{M_V}\left(
  \begin{array}{c}
  \vec{p}\cdot\vec{\varepsilon'}_\lambda\\
  \vec{p}\frac{\vec{p}\cdot\vec{\varepsilon'}_\lambda}{p^0+M_V}+M_V\vec{\varepsilon'}_\lambda
  \end{array}\right).
\end{eqnarray}

If the polarization axis of vector meson is taken along the
direction of its three-momentum $\vec{p}$,
 i.e. the helicity state, it is clear that $\vec{\varepsilon '}_0$ is
parallel with $\vec{ p}$. So we can get $\vec{\varepsilon '}_0\times \vec{p}=0$.
Furthermore, due to the relation $\vec{\varepsilon' }_{
\lambda}\cdot\vec{\varepsilon '}_{\lambda'}=\delta_{\lambda \lambda'}$,
we also have $\vec{\varepsilon '}_{\pm1}\cdot\vec{ p}=0$. And then the
helicity state of the vector meson can be rewritten as
\begin{eqnarray}
\varepsilon_{\pm1}=\left(
  \begin{array}{c}
  0\\
  \vec{\epsilon'}_{\pm1}
  \end{array}\right),
\varepsilon_{0}=\frac{1}{M_V}\left(
  \begin{array}{c}
  |\vec{p}|\\
  p^0\vec{\epsilon'}_{0}
  \end{array}\right)\label{epV}
\end{eqnarray}

\section{PROPERTIES of THE VPP AND VVP VERTICES}\label{VP}
For the VPP vertex, we have the interaction Lagrangian
\begin{eqnarray}
\CL_{VPP}=g_{VPP} V_\mu(P_1\partial^\mu P_2-P_2\partial^\mu P_1),
\end{eqnarray}
and the corresponding vertex function is
\begin{eqnarray}
-it_{V\to PP}&&=g_{VPP} \varepsilon_{\mu}(p_1-p_2)^\mu\nonumber\\&&
=g_{VPP}\varepsilon_{\mu}[p_1-(p_V-p_1)]^\mu\nonumber\\&&
=2g_{VPP}\varepsilon\cdot p_1
\end{eqnarray}
where $p_1$, $p_2$ and $p_V$ represent momenta of the two
pseudoscalar mesons and the vector meson V. If $\vec{p}_1$ and
$\vec{p}_V$ are collinear in some reference frame, according
to Eq.(\ref{epV}) the helicity states of the vector meson
have the property
\begin{eqnarray}
\varepsilon_{\pm1}\cdot p_{1}=
  0\cdot p_1^0-\vec{\varepsilon'}_{\pm1}\cdot \vec{p_1}=0.
\end{eqnarray}
Therefore in the reference frame where the three momenta $p_1$
and $p_V$ are parallel, the VPP vertex has the property that the
produced V meson can only be in the helicity state with $\lambda=0$,
which is a result of angular momentum conservation.

For the VVP vertex, the Lagrangian is written as
\begin{eqnarray}
\CL_{VVP}=g_{VVP}\varepsilon^{\mu\nu\alpha\beta}\partial_\mu
V_{1\nu}
\partial_\alpha V_{2\beta}.
\end{eqnarray}

The corresponding vertex function is
\begin{eqnarray}
-it_{V_1\to V_2P}=g_{VVP}\varepsilon_{\mu\nu\alpha\beta}p^{\mu}
\varepsilon_1^{\nu} q^{\alpha}\varepsilon_2^{*\beta}
\end{eqnarray}
with $p$ and $q$ denoting the momentum of the vector mesons $V_1$
and $V_2$. $\varepsilon_1$ and $\varepsilon_2$ represent their
corresponding polarization vectors. We can rewrite the vertex
function in the following form:
\begin{eqnarray}
-it_{\smsm{V_1\to V_2P}}&&=g_{\smsm{VVP}}[p^{0}
(\vec{\varepsilon}_1\times \vec{q}
)\cdot\vec{\varepsilon}_2^*-\varepsilon_1^{0}(\vec{p}\times
\vec{q})\\&&\nonumber \cdot \vec{\varepsilon}_2^*+q^{0}(\vec{p}\times
\vec{\varepsilon}_1)\cdot
\vec{\varepsilon}^*_2-\varepsilon^{*0}_2(\vec{p}\times
\vec{\varepsilon}_1)\cdot \vec{q}].
\end{eqnarray}
This expression shows that, if the three-momenta of the three
particles are collinear and the helicity of $V_2$ is $0$, the
vertex function should vanish due to the equations $\vec{p}\times\vec{q}=\vec{\varepsilon}_{2,\lambda=0}\times \vec{q}
=\vec{\varepsilon}_{2,\lambda=0}\times \vec{p}=0$ (see Appendix \ref{POV}).
It means only the helicity states with $\lambda=\pm 1$ contribute,
and the produced vector meson $V_2$ should have $\rho_{00}=0$. When
the vector meson $V_1$ has a vanishing momentum or the calculation is
performed in its rest frame, similar arguments also hold. This property
can be understood in the following way. Let us consider the process
$V_1$ decaying to $V_2$ and P. In this case, due to the conservation
of parity and angular momentum, the orbital angular momentum of the
final two particles can only be 1. If we choose the z axis along the
momentum of $V_2$ in the $V_1$'s rest frame, the magnetic quantum
number of the initial state $V_1$ (denoted as m) can only have the
same value as the helicity of the $V_2$ (denoted as $\lambda_2$) due
to the conservation of z-component of total angular momentum. In this
case $\lambda_2=0$ is forbidden, because the coupling of the spin
states of $V_1$ and $V_2$ ($|1,0>$ and $|1,0>$) with orbital angular
momentum state $|1,0>$ is vanishing due to the Clebsch-Gordan
coefficient $<10,10|10>=0$.

\end{appendix}

\end{document}